\documentclass[preprint,aps,prl,amssymb,nobibnotes,letterpaper,superscriptaddress]{revtex4-1}

\usepackage{graphicx}
\usepackage[utf8]{inputenc}
\usepackage{hyperref}% use for hypertext links, including those to external documents and URLs

\begin{document}
	
	\newcommand{\PSI}{Swiss Light Source, Paul Scherrer Institute, 5232 Villigen-PSI, Switzerland }	
	\newcommand{\SwissFEL}{SwissFEL, Paul Scherrer Institute, 5232 Villigen-PSI, Switzerland }
	\newcommand{\ETHTh}{Materials Theory, ETH Z\"urich, 8093 Z\"urich, Switzerland}
	\newcommand{\MPI}{Max Planck Institute for the Structure and Dynamics of Matter, CFEL, 22761 Hamburg, Germany } \newcommand{\ETHq}{Institute for Quantum Electronics, ETH Z\"urich, 8093 Z\"urich, Switzerland }
	\newcommand{\FHIpc}{Department of Physical Chemistry, Fritz Haber Institute of the Max Planck Society, 14195 Berlin, Germany}
	\newcommand{\FHIic}{Department of Inorganic Chemistry, Fritz Haber Institute of the Max Planck Society, 14195 Berlin, Germany}
	\newcommand{\LCLS}{LCLS, SLAC National Accelerator Laboratory, Menlo Park, California 94025, USA}
	\newcommand{\UR}{Department of Physics, University of Regensburg, 93040 Regensburg, Germany}
	%\preprint{}
	
	\setlength{\skip\footins}{4.5mm}
	\renewcommand{\footnoterule}{}
	%PRL
	%\title{Ultrafast transient enhancement of the structural order parameter in EuTiO$_3$}

	\title{Ultrafast transient increase of oxygen octahedral rotations in a perovskite}

	\author{M. \surname{Porer}}
	\email[]{Michael@Porer.org}
	\affiliation{\PSI}
	
	%\author{Michael \surname{Fechner}}
	\author{M. \surname{Fechner}}
	\affiliation{\MPI}

	\author{\mbox{M. \surname{Kubli}}}
	\affiliation{\ETHq}
	
	\author{M. J. \surname{Neugebauer}}
	\affiliation{\ETHq}
	
	\author{S. \surname{Parchenko}}
	\affiliation{\PSI}
	
	\author{V. \surname{Esposito}}
	\affiliation{\SwissFEL}

	\author{A. \surname{Narayan}}
	\affiliation{\ETHTh}
	
	\author{N. A. \surname{Spaldin}}
	\affiliation{\ETHTh}
	
	\author{R. \surname{Huber}}
	\affiliation{\UR}
	
	\author{M. \surname{Radovic}}
	\affiliation{\PSI}
	
	\author{E. M. \surname{Bothschafter}}
	\affiliation{\PSI}

	\author{J. M. Glownia}
	\affiliation{\LCLS}
	
	\author{T. \surname{Sato}}
	\affiliation{\LCLS}
	
	\author{S. \surname{Song}}
	\affiliation{\LCLS}
	
	\author{S. L. \surname{Johnson}}
	\affiliation{\ETHq}	\affiliation{\SwissFEL}
	\author{U. \surname{Staub}}
	\email[]{Urs.Staub@psi.ch}
	\affiliation{\PSI}
	
	\date{\today}

	%Abstract PRL style
	%We monitor the order parameter of the soft-mode driven structural lattice distortion in EuTiO$_3$ via femtosecond hard X-ray diffraction after ultrafast photoexcitation across the band gap. Within an initial sub-ps time window, the order parameter given by the antiferrodistortive rotation of the oxygen octahedra is transiently increased. A relaxation below the initial baseline follows on a few ps timescale. Nonequilibrium frozen phonon DFT calculations reproduce an increase of the optimal rotation amplitude with the photodoping level for simultaneous doping of electrons and holes. The subsequent relaxation is explained by a combination of carrier recombination and heating of the crystal lattice.
	
	\begin{abstract}
		The ability to control the structure of a crystalline solid on ultrafast timescales bears enormous potential for information storage and manipulation or generating new functional states of matter \cite{buzzi2018}. In many materials where the ultrafast control of crystalline structures has been explored, optical excitation pushes materials towards their less ordered high temperature phase \cite{eichberger2010, koopmans2010, rohwer2011, porer2014, beaud2014, esposito2017, porer2018, fausti2011} as electronically driven ordered phases melt and possible concomitant structural modifications relax. Nonetheless, for a few select materials it has been shown that photoexcitation can slightly enhance the amplitude of an electronic ordering phenomenon (i.e. its electronic order parameter) \cite{fausti2011, kim2012, matsubara2015, singer2016, mitrano2016}. Here we show via femtosecond hard X-ray diffraction that photodoping of the perovskite EuTiO$_3$ transiently increases the order parameter associated with a purely structural \cite{ellis2012} phase transition represented by the antiferrodistortive rotation of the oxygen octahedra. This can be understood from an ultrafast charge-transfer induced reduction of the Goldschmidt tolerance factor \cite{goldschmidt1926}, which is a fundamental control parameter for the properties of perovskites. 
	\end{abstract}
	
	%	\pacs{63.20.K-,64.60.-i}
	%\keywords{}
	
	%A prominent example is the enhancement of unconventional superconductivity via ultrafast suppression of a competing ordering phenomenon such as stripe order \cite{Fausti2011} or charge-density waves (CDWs) \cite{find UP paper}. For strongly correlated systems such enhancements of one oder-parameter against another enable a deeper understanding of the interplay between the underlying ordering phenomena.
	
	\maketitle

	Phase transitions are commonly described by an order parameter (OP) that represents the degree of order in the system. Within the ordered phase, the OP quantifies the amplitude of an ordering phenomenon or the distance to the critical point with respect to a tuning parameter such as temperature, structural parameters or the electronic chemical potential. Examples of OPs include the amplitude of a symmetry-breaking distortion of a crystal structure and the amount of charge carriers in a charge/orbital-density wave or a superconducting condensate. In correlated materials, structural or electronic ordering phenomena often occur simultaneously or compete with each other \cite{morosan2012}. Slight changes of a tuning parameter can induce a phase transition with potentially useful functionality. Optical control of such tuning parameters opens up the possibility to manipulate an OP and thereby control possibly connected macroscopic material properties on ultrafast timescales. Novel routes to optically increase an OP have been explored recently with the goal of enhancing or inducing functional properties. Examples include light enhanced superconductivity \cite{mitrano2016,fausti2011}, increased charge \cite{singer2016} or spin \cite{kim2012} density wave amplitudes. Significant potential exists for increasing an order parameter by coherently controlling the structural parameters \cite{mankowsky2014,subedi2014,mankowsky2017,kozina2019}. 
	\par

	Second-order purely structural phase transitions are driven by anharmonic interactions within the phonon system \cite{cowley1980,dove1997}. When a tuning parameter reaches the critical point, the structure spontaneously distorts along a vibrational coordinate which lowers the symmetry of the crystal. The corresponding OP is the mean displacement amplitude \cite{dove1997}. At the critical point of a displacive \cite{dove1997} structural transition, the eigenfrequency of the respective vibrational mode, called softmode, approaches zero and the susceptibility of the OP with respect to the tuning parameter diverges.

	\par
	EuTiO$_3$ is a cubic perovskite (Pm$\overline{3}$m) at high temperature which undergoes a structural phase transition around $T \approx 290 K$ driven by an acoustic softmode at the $R$-point which manifests as an antiferrodistortive (AFD) rotation of the oxygen octahedra (Fig. 1a) , very similar to that in SrTiO$_3$ \cite{ellis2012,goian2012,bettis2011}. The rotation angle $\varphi$ is the order parameter of the transition \cite{dove1997}.
	Here we photoexcite EuTiO$_3$ in its low temperature structurally distorted phase (I4/mcm) via ultrafast optical excitation across its direct band gap of ~0.9 eV \cite{lee2009} (see Methods). Via subsequent femtosecond hard X-ray pulses we monitor the intensities of AFD induced X-ray superlattice (SL) reflections (Fig. 1b) which are proportional to $\varphi^2$ \cite{porer2018}.

	\par
	
	\begin{figure*}[t]
		\includegraphics[width=0.9\textwidth]{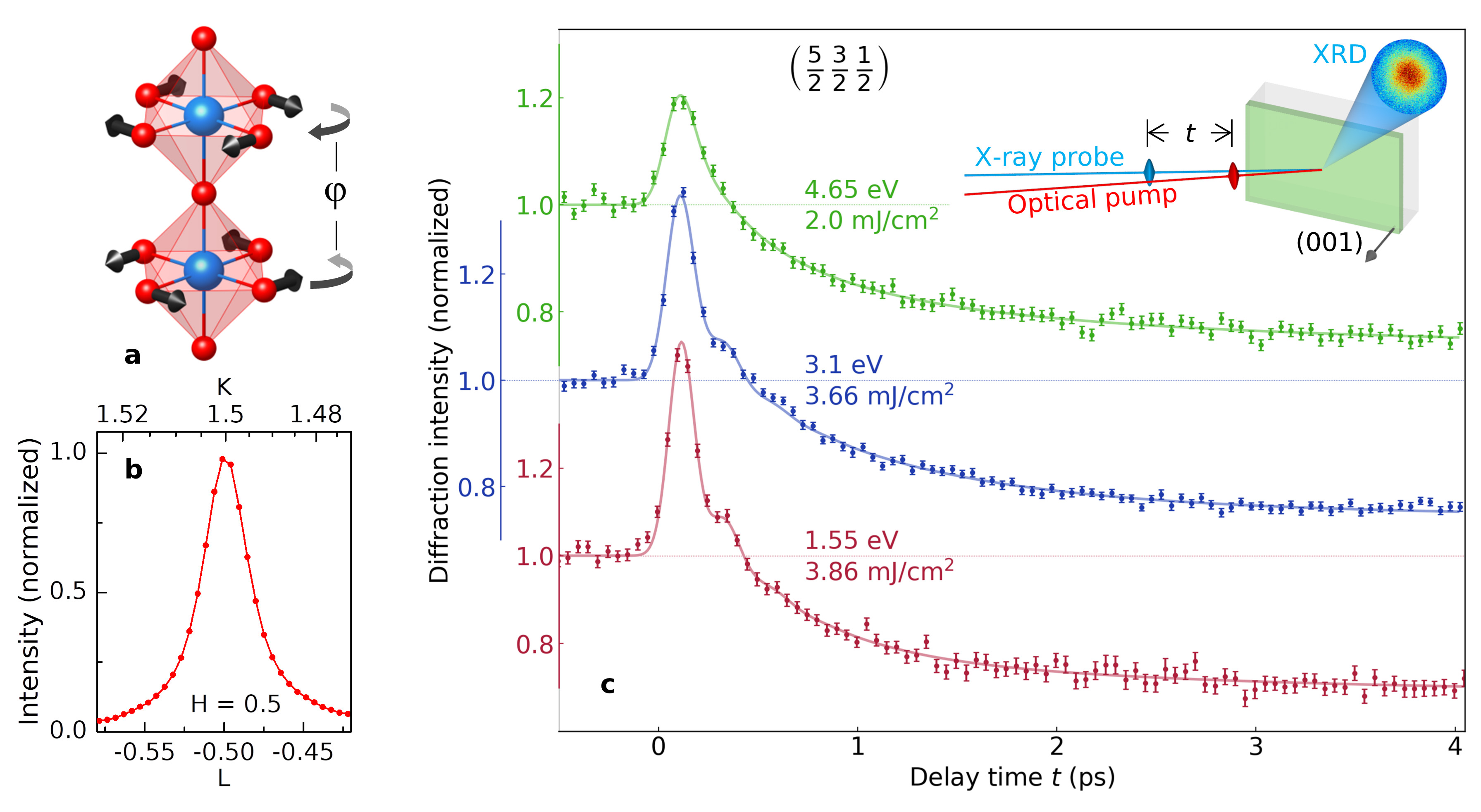}
		\caption{ (a) Ti (blue) and O (red) atoms of two cubic unit cells stacked along the $c$-axis. Arrows indicate atomic motions of the oxygen sites during the AFD transition and the direction of the octahedral rotation angle $\varphi$. (b) Equilibrium rocking curve of a typical superlattice reflection of the 40 nm thin EuTiO$_3$ film measured at $T = 120\,\mathrm{K}$. (c) Dynamics of the (5/2 3/2 1/2) superlattice reflection intensity upon photoexcitation at $T = 120\,\mathrm{K}$. The labels denote the pump photon energy and the absorbed excitation fluence. The blue and red solid lines result from a fit of the phenomenological model explained in the text and the Supplementary Information. The green solid line is a guide to the eye. Inset: Schematic of the femtosecond X-ray diffraction (XRD) experiment.
			\label{fig1}}
	\end{figure*}

	\par
	Figure 1c shows the normalised intensity of the \mbox{(5/2 3/2 1/2)} SL reflection as a function of the pump-probe delay for a series of excitation conditions. The green/blue/red data points show the transient SL intensity obtained for excitation with photon energies of 4.65/3.1/1.55 eV for an absorbed fluence of 2.0/3.66/3.86 mJ/cm$^2$ injecting a density (see Methods) of 0.04/0.11/0.23 electron-hole (eh) pairs per formula unit (FU), respectively.
	Independent of the pump photon energies and fluences/injected carrier densities, we observe a transient increase of the superlattice diffraction intensity ($I\textsubscript{AFD}$) within an initial sub-ps time window. The enhancement is similar for another AFD-induced SL reflection with different X-ray momentum transfer (Supplementary Figure S2). Quantitatively, the maximum observed enhancement factor of $I\textsubscript{AFD}$ is 1.4 for the highest injected density with 1.55 eV photons. The subsequent decay of $I\textsubscript{AFD}$ intersects the initial baseline. For excitation with photon energies below 4.65 eV the decay exhibits a superimposed strongly damped oscillatory component with a frequency of approximately $4\,\mathrm{THz}$. Furthermore, when lowering the excitation intensity we find that the decay slows down and the lifetime of the enhanced diffraction intensity increases (Fig. 2). The base temperature does not significantly influence the dynamics (Fig. S3). As $I\textsubscript{AFD}$ scales with $\varphi^2$, the enhanced diffraction efficiency may originate from an increase of the OP. 
	\par	
	To describe this behavior we calculate the total energy within the I4/mcm unit cell at a series of fixed rotation angles (Fig. S4) and from there we derive the double-well potential of the AFD mode (SI). Fig. 3b shows the resulting potential for various doping levels. We obtain a rotation angle for zero doping of $\varphi_0=6.3^\circ$. With increasing $\rho$, we find both a deepening of the potential and, importantly, an increasing displacement of the potential minimum from $\varphi_0$. The latter yields a driving force towards an increased AFD rotation which can explain the increase of SL diffraction intensity during the initial time-window (Fig. 1c). An increased excess energy of the doped carriers, implemented via an elevated electronic temperature, is predicted to further enhance the effect (Fig. S5). This trend agrees with our experimental observation that fewer eh pairs are needed to generate equivalent enhancements at higher excitation frequencies, e.g. $0.11$ eh pairs per FU generated by 3.1 eV pump photons yield a similar enhancement as $0.23$ eh pairs per FU injected by 1.55 eV photons (Fig. 1c, blue and red datapoints).
	\par
	
	In principle, deepening of the soft-mode potential with constant $\varphi$ could also enhance the reflection intensity by reducing the average incoherent thermal displacements of the local rotations. This would increase the Debye-Waller factor (DWF) and consequently the reflection efficiency. For our experimental conditions, we expect an initial DWF of $>0.99$ for the \mbox{(5/2 3/2 1/2)} SL reflection based on the thermal displacement parameters of EuTiO$_3$ \cite{allieta2012}. This rules out an increase of $I\textsubscript{AFD}$ by more than 1\% from changes in the DWF. Furthermore, the scaling of the DWF with the X-ray momentum transfer $\mathbf{q}$ implies that the reflection efficiency would be more strongly enhanced at larger $\mathbf{q}$, which is not observed (Fig. S2). Both arguments exclude a significant influence of photoinduced deepening of the double-well potential on the observed dynamics of $I\textsubscript{AFD}$. 
	
	\begin{figure}[t]
		\includegraphics[width=0.5\columnwidth]{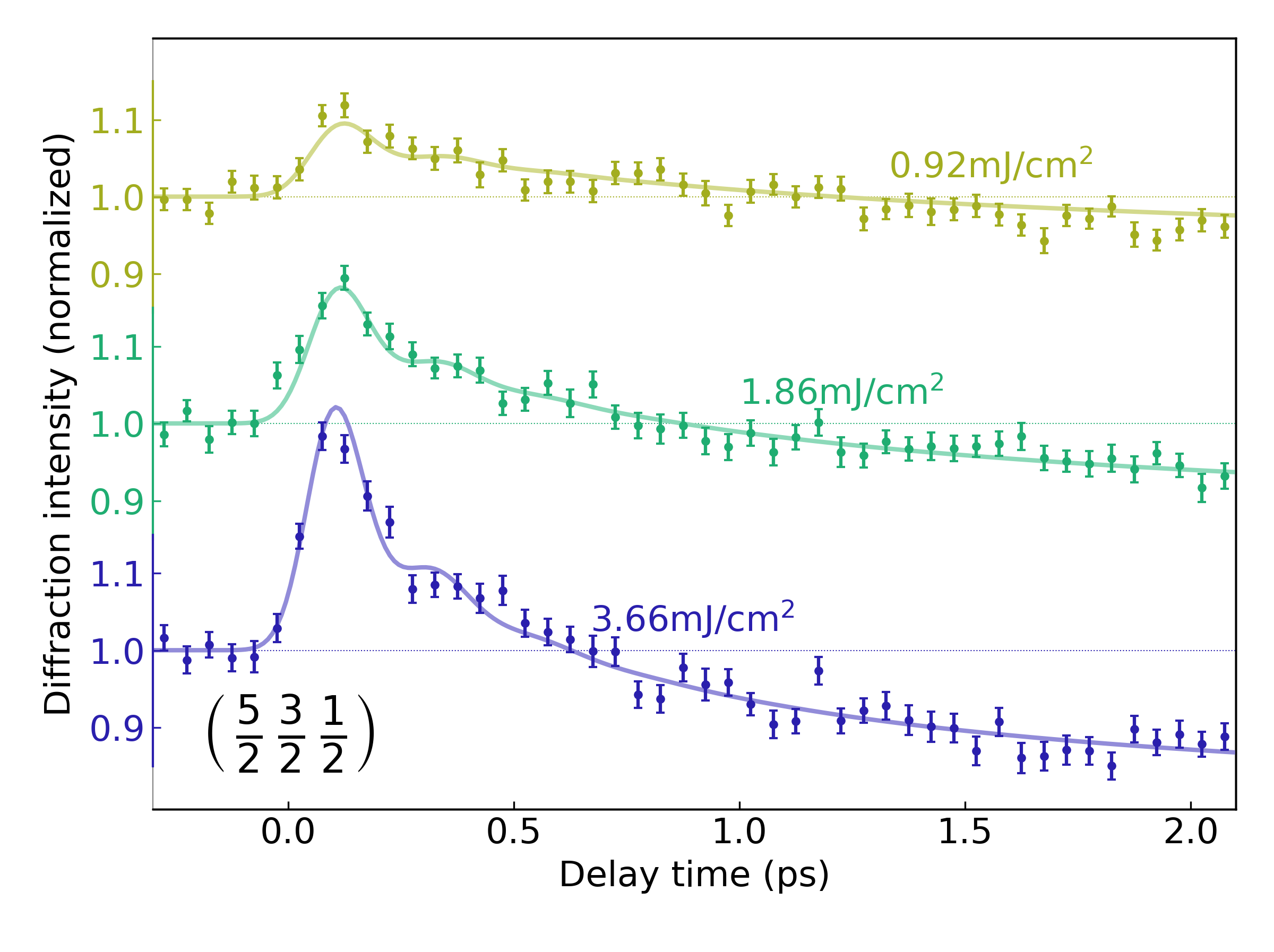}
		\caption{Transient superlattice diffraction intensity for a series of excitation fluences. The pump photon energy is set to $3.1\,\mathrm{eV}$ and the temperature is $120\,\mathrm{K}$. The solid lines are derived from a numerical fit to the model described in the text.\label{fig2}}
	\end{figure}	
	\begin{figure}[t]
		\includegraphics[width=0.5\columnwidth]{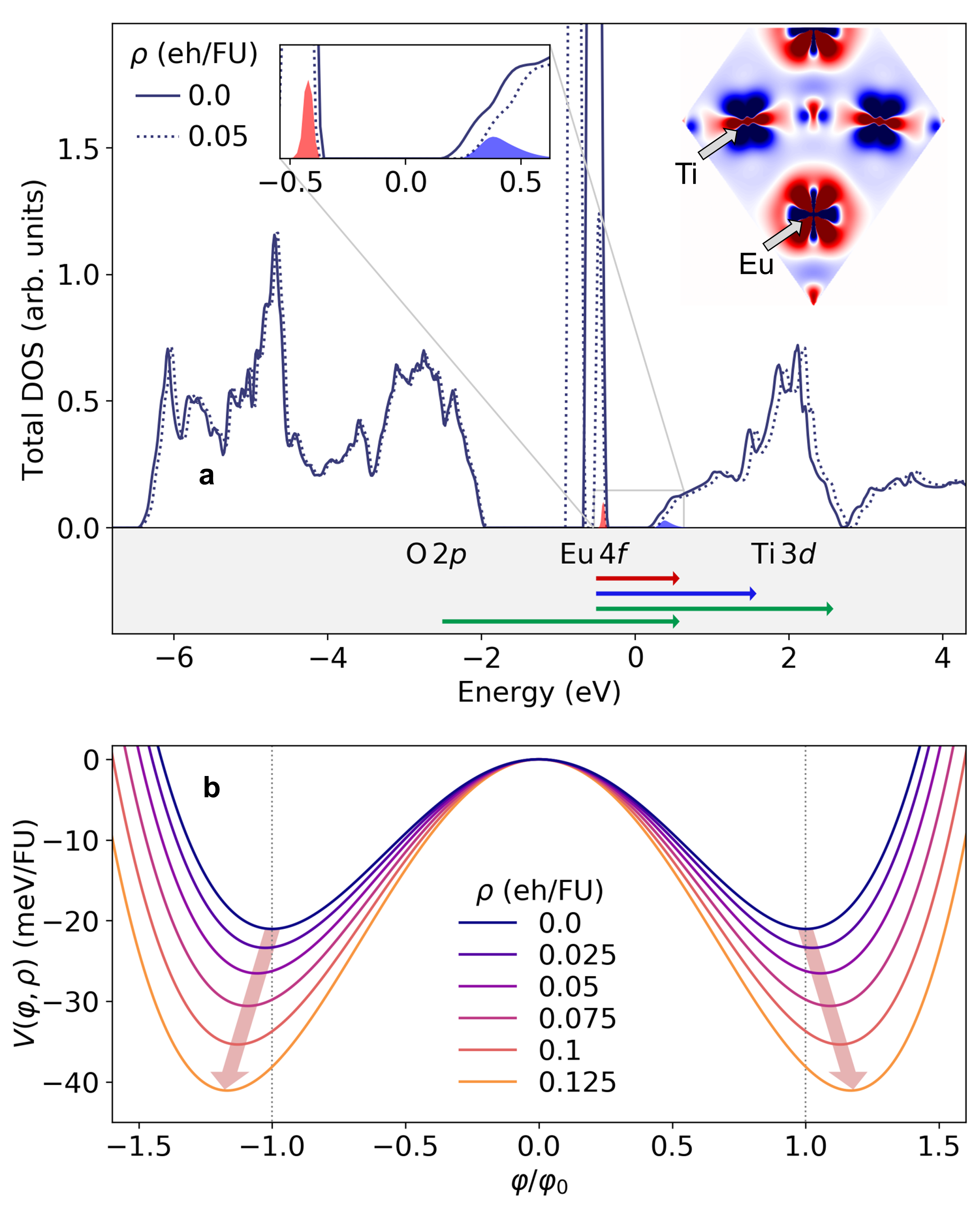}
		\caption{(a) Calculated EuTiO$_3$ total density of states (DOS) in equilibrium and for an eh concentration of 0.05 eh/FU (dotted line). Zero is set to the bottom of the conduction band. Arrows indicate the charge-transfer excitations for photon energies of 1.55 eV (red), 3.1 eV (blue), 4.65 eV (green.) Right inset: Slice in the (101) plane of the eh induced total e$^{-}$ density difference (blue/red: increase/decrease of electron density). Left inset: Zoom of the DOS in the vicinity of the Fermi level. The filled areas show the density of doped holes (red) and electrons (blue). (b) Double well potential of the AFD soft mode calculated for a series of eh concentrations $\rho$.  
			\label{fig3}}
	\end{figure}
	
	\par While the observed photoinduced increase of the OP is contrary to its evolution for increasing temperature (and phonon entropy) in thermal equilibrium, it does however not imply an actual decrease of phonon entropy. The increase of $\varphi$ can be understood intuitively in terms of a reduction of the Goldschmidt tolerance factor \cite{goldschmidt1926}: Photoexcitation across the Eu $4f$ - Ti $3d$ charge-transfer gap of EuTiO$_3$ (indicated by the top three arrows at the bottom of Fig. 3a) removes $f$-electrons from the Eu$^{2+}$ A-site ions and shrinks the surrounding electron clouds. Analogously, the Ti$^{4+}$ B-site ions expand due to their additional electronic charge. Both size effects decrease the tolerance factor, which is known to cause a more distorted perovskite structure in the static case. We illustrate this in Fig. 3a right inset, the calculated change in charge density in the (101) plane (cubic notation). We see that the electron density decreases approximately spherically at the Eu sites and increases at the Ti sites. This unusual decrease in electron density on the Eu$^{2+}$ ion on photodoping is the origin of the stark contrast in behavior between EuTiO$_3$ and SrTiO$_3$ on photoexcitation, in spite of their similar crystallographic structures and O $2p$ -- Ti $3d$ electronic bands \cite{bettis2011,birol2013}. In the case of EuTiO$_3$, the largely unhybridized Eu $4f$ states form an additional set of occupied bands above the oxygen $2p$ valence band; these localized $4f$ states are electron depleted by charge-transfer excitation reducing the size of the Eu$^{2+}$ cation. In contrast, charge transfer excitation across the O $2p$ -Ti $3d$ band gap of SrTiO$_3$ {\it reduces} the octahedral distortion \cite{porer2018}.

	\par
	
	Following the excitation, both phonon-mediated eh recombination and cooling of the electronic system via electron-phonon scattering reduce the electronically driven enhancement of $\varphi$ and heat the phonon system. A rapidly increased temperature of the vibrational system can be expected to further relax the distortion via anharmonic phonon-phonon interactions, similar to a thermally driven phase transition.
	\par
	 All optical experiments on a EuTiO$_3$ film (Methods, SI) show a decay dynamics of the photoinduced reflectivity change for 1.55 eV probe photons that resembles a bimolecular decay dynamics. Similar dynamics is observed when probed in the multi-THz regime. We ascribe this decay to an excitation-dependent carrier recombination process (SI). 
	 \par
	To model the dynamics of chemical potential $\varphi$, we introduce a time dependency of chemical potential $V(\varphi,\rho)$ via a transient doping level $\rho(t)$. The onset dynamics of $\rho(t)$ reflects the pump pulse duration and injected number of eh pairs and the decay dynamics accounts for the recombination process (SI). The temperature increase of the lattice is included via an additional relaxation of $V(\varphi,\rho)$ towards the high temperature phase along its dependency on $\rho$. We scale this additional component phenomenologically with the amount of recombined charge carriers at time $t$ (SI). Solving the equation of motion for a time-dependent potential $V(\varphi(t),\rho(t))$ yields $\varphi(t)$ \cite{porer2018} (SI) which we numerically fit via $\varphi(t)^2/\varphi_0(t)^2$ to the experimental data (solid lines, Fig. 1c (red and blue curves), and Fig. 2).
	
	\par
	The model reproduces the slow down of the decay dynamics for lowering the excitation fluence. As the slow down is described by an eh density-dependent carrier recombination, we conclude that the lifetime of the enhanced octahedral rotation is mainly governed by the lifetime of the photoinjected carriers. The weak oscillatory component on the decay dynamics of $I\textsubscript{AFD}$  (Fig. 1 (c), red and blue curves) is well reproduced by a model consiting of a displacive excitation caused by a shift of the potential minimum, with the correct phase and frequency and thus further supports a displacively driven structural dynamics initiated by the optical excitation. The damping of the soft mode is fixed to yield a coherence lifetime of $0.14\,\mathrm{ps}$. The absence of an oscillation after excitation with 4.65 eV photons is possibly related to a delayed build-up of the displacive force due to the opposite effects of the simultaneous O$2p$-Ti$3d$ and Eu$4f$-Ti$3d$ charge transfer excitations.
	\par

	Ultrafast control of a structural distortion in the opposite direction to its thermal transition by optical excitation opens up new possibilities for ultrafast tuning of electronic and magnetic properties \cite{imada1998,aken2004,varignon2017}.
	In particular, the ability to tune rotations of oxygen octhahedra, allows in turn the tuning of magnetic and electronic properties that are strongly correlated with the tolerance factor and structural distortions, such as magnetic orderings or electric polarization, on ultrafast timescales. 
	
	%As theoretically investigated for SrTiO$_3$ \cite{porer2018}, a continuous change of the photodoping level can yield a transition from a double- to a single-well potential. A thereto reversed dependency of the soft-mode potential on the doping level, as we demonstrate for EuTiO$_3$, might thus be employed to induce a symmetry break of a single-well soft-mode potential for suitable equilibrium conditions. This scenario is expected similarly for ultrafast resonant tuning of structural parameters \cite{subedi2014}. Reduced excitation intensities and controlled photon excess energies diminish the amount of thermal heating and can extend the lifetime of the doped carriers, which might be sufficient to tip the balance for inducing such the transition close to a critical point. Even without directly enabling a structural symmetry break, one may consider to induce a transition that is tuned by the distortion amplitude itself. Our study points out a novel route for controlling select structural soft-mode potentials to drive (connected) phase transitions with potentially functional modifications of macroscopic material properties. 

	\section{Methods}
	EuTiO$_3$ films were grown by Pulsed Laser Deposition (PLD) at the PLD facility of the Surface/Interface Spectroscopy (SIS, X09LA) beamline at the Swiss Light Source (SLS) - Paul Scherrer Institute. The films were deposited on commercial SrTiO$_3$ (001) substrates (purchased from SurfaceNet GmbH) in a very low oxygen partial pressure of $10^{-7}$ mbar (base pressure of the PLD chamber $10^{-9}$ mbar) and at a temperature of $750\,\mathrm{^\circ C}$. The phase transition of the 40 nm film was characterised using static X-ray diffraction (Figures 1b and S1). Compared to EuTiO$_3$ ceramic material \cite{goian2012}, we find the critical temperature to be elevated to $T_c \approx 650\, \mathrm{K}$ which we attribute to oxygen defects \cite{goian2012} and/or residual strain \cite{ryan2013}. \par
	For time resolved X-ray diffraction experiments, we photoexcite the EuTiO$_3$ thin film with ~50 fs long pulses derived either from the fundamental \mbox{($\left<E_p\right>=1.55\,\mathrm{eV}$)}, the second harmonic \mbox{($\left<E_p\right>=3.1\,\mathrm{eV}$)} or the third harmonic \mbox{($\left<E_p\right>=4.65\,\mathrm{eV}$)} of a Ti:Sapphire amplifier system. We estimate the absorbed fluence in the film via an optical transfer matrix formalism \cite{pettersson1999} based on the the optical constants of EuTiO$_3$ and the SrTiO$_3$ substrate (SI). The injected densities of eh pairs for the laser harmonics are $6\cdot10^{-2}$, $3\cdot10^{-2}$ and $2\cdot10^{-2}$ per formula unit and 1 mJ/cm$^2$ absorbed fluence of 1.55 eV, 3.1 eV and 4.65 eV excitation energies, respectively. 
	\par
	Approximately 50 fs long X-ray pulses with 9.5 keV photon energy delivered by LCLS (SLAC) were used to probe the SL reflection intensity after excitation (SI). The sample was cooled with a N$_2$ cryo-blower to $T=120\,\mathrm{K}$ to remain above the critical temperature for the AFD transition of the SrTiO$_3$ substrate at $T=105\,\mathrm{K}$ \cite{bettis2011}.\par
	Supplementary optical pump-probe studies were performed at the FEMTO-facility at Swiss Light Source using ~80 fs long pulses with a central photon energy of 1.55 eV. Additional time resolved optical-pump/multi-THz probe studies were performed at the University of Regensburg (SI).
	
	The density functional theory (DFT) computations are carried out utilizing the all-electron full-potential linearised augmented-plane wave implementation within the elk code \cite{elk}. As an approximation for the exchange-correlation functional we apply a generalised gradient approach plus U (GGA+U) and apply a $U = 6.2\,\mathrm{eV}$ and $J = 1\,\mathrm{eV}$, respectively on the Eu $4f$-states. The muffin tin radii for Eu, Ti and O are set to $1.3\,\mathrm{\AA}$, $0.9\,\mathrm{\AA}$ and $0.75\,\mathrm{\AA}$. We used a basis set of $l_{lmax_\mathrm{APW}}
=10$, $8\times 8 \times 6$  k-point sampling of the Brillouin zone and took the product of the average muffin tin radius and the maximum reciprocal lattice vector to be 8.5. Further numerical parameters are adjusted to the high-quality ('\texttt{highq .true.}') settings within the Elk code.

	\section{Acknowledgements}
	We thank H. Lemke for support in processing data acquired at LCLS. Static X-ray characterization of the sample was performed at the X04SA beamline (SLS) with technical assistance from D. Meister. We thank J. L. Mardegan, J. Raab and M. Furthmeier for technical assistance on the all-optical studies. The research leading to these results has received funding from the Swiss National Science Foundation and its National Centers of Competence in Research, NCCR MUST and NCCR MARVEL. E.M.B. acknowledges funding from the European Community’s Seventh Framework Programme (FP7/2007-2013) under grant agreement No. 290605 (PSI-FELLOW/COFUND). Computing resources were provided by the MERLIN cluster at the Paul Scherrer Institute. 
	\section{Author contributions}
	
	M. P., S. L. J. and U. S. conceived the experiment. M. P., M. K., M. J. N., S. P., S. L. J. and U. S. performed the time-resolved diffraction experiment together with J. M. G, S. S. and T. S.. M. R. grew the sample. M. P., M. J. N., M. K., and S. P. analysed FEL data. M. P. performed supplementary time-resolved optical experiments. M. P., S. L. J, M. F. and U. S. interpreted the data with support from N. A. S.. E. M. B., V. E., M. K., M. J. N. and M. P. characterised the static diffraction properties of sample candidates. M. P. implemented the DFT constraints in the \emph{elk}-package and performed calculations in collaboration with M. F., support from A. N. and input from N. A. S.. A. N. performed additional DFT calculations. M. P., M. F., S. L. J. and U. S. wrote the paper and all authors contributed to the final version.
	%\bibliographystyle{unsrt}
	%\bibliography{ETO}

{}

\end{document}